\documentclass[conference]{IEEEtran}
\IEEEoverridecommandlockouts

\usepackage{academicons}

\usepackage{cite}
\usepackage{graphicx}
\usepackage{comment}
\usepackage{balance}
\usepackage{acronym}
\usepackage{amsmath}
\usepackage{tabularx}
\usepackage[usenames,dvipsnames]{xcolor}
\usepackage{subcaption}
\usepackage{booktabs}
\usepackage{multirow}

\def\BibTeX{{\rm B\kern-.05em{\sc i\kern-.025em b}\kern-.08em
    T\kern-.1667em\lower.7ex\hbox{E}\kern-.125emX}}
    
\title{Concurrent Speaker Detection: A multi-microphone Transformer-Based Approach\thanks{This project has received funding from the European Union’s Horizon 2020 Research and Innovation Programme, Grant Agreement No. 871245, and was also supported in part by the Israeli Council for Higher Education, Data Science Program (``Audience" project), and by the Israeli Ministry of Science \& Technology.}
}

\author{\IEEEauthorblockN{Amit Eliav}
\IEEEauthorblockA{\textit{Faculty of Engineering} \\
\textit{Bar-Ilan University},
Ramat-Gam, Israel \\
amiteli@biu.ac.il}
\and
\IEEEauthorblockN{Sharon Gannot}
\IEEEauthorblockA{\textit{Faculty of Engineering} \\
\textit{Bar-Ilan University},
Ramat-Gam, Israel \\
sharon.gannot@biu.ac.il
}
}

\acrodef{STFT}{Short-Time Fourier Transform}
\acrodef{ISTFT}{inverse short-time Fourier transform}
\acrodef{TF}{Time-Frequency}
\acrodef{RTF}{Relative Transfer Function }

\acrodef{LCMV}{Linearly Constrained Minimum Variance}

\acrodef{CSD}{Concurrent Speaker Detection}
\acrodef{VAD}{Voice Activity Detection}
\acrodef{OSD}{Overlapped Speech Detection}

\acrodef{NLP}{Natural Language Processing}

\acrodef{LSTM}{Long Short-Term Memory}
\acrodef{BLSTM}{Bidirectional Long Short-Term Memory}
\acrodef{RNN}{Recurrent Neural Networks}
\acrodef{CNN}{Convolutional Neural Networks}

\acrodef{GMM}{Gaussian mixture model}

\acrodef{ViT}{Vision Transformer}
\acrodef{CLS}{Class Token}
\acrodef{MHA}{multi-head attention}
\acrodef{TCN}{Temporal Convolutional Networks}

\acrodef{CE}{Cross-Entropy}
\acrodef{CS}{Cost-Sensitive}
\acrodef{LS}{Label-Smoothing}

\acrodef{mAP}{mean average precision}

\acrodef{SOTA}{state-of-the-art}
\acrodef{AST}{Audio Spectrogram Transformer}
\acrodef{ViT}{Vision Transformer}

\begin{document}
\bstctlcite{IEEEexample:BSTcontrol}


\maketitle

\begin{abstract}
\label{abstract}
We present a deep-learning approach for the task of \ac{CSD} using a modified transformer model. Our model is designed to handle multi-microphone data but can also work in the single-microphone case. The method can classify audio segments into one of three classes: 1) no speech activity (noise only), 2) only a single speaker is active, and 3) more than one speaker is active. We incorporate a \ac{CS} loss and a confidence calibration to the training procedure. The approach is evaluated using three real-world databases: AMI, AliMeeting, and CHiME~5, demonstrating an improvement over existing approaches.
\end{abstract}

\section{Introduction}
\label{Introduction}
Speaker detection, namely the ability to identify and track the activities of individual speakers in an audio stream, is an important task with many practical applications. In particular, \acf{CSD} is the problem of identifying speakers' presence and overlapping activity in a given audio signal. It classifies audio segments into three classes, namely: 1) no speech activity (noise only), 2) only a single speaker is active, and 3) more than one speaker is active. 
A reliable \ac{CSD} is a key component in audio scene analysis and speech processing applications, e.g., speech detection, speaker counting and diarization, and multi-microphone spatial processing in ``cocktail party'' scenarios.
\ac{CSD} is a challenging task due to the complex nature of human speech. Accent, pitch, and speaking style variations can make the identification and detection of the speakers' activity challenging. Consequently, developing effective \ac{CSD} approaches is an active area of research, aiming to improve accuracy and robustness.

In \cite{8553564}, a multichannel \ac{CSD}, based on \ac{CNN} architecture, was used as a building block of a \ac{LCMV} beamformer for controlling the estimation of its components, based on the speakers' activity patterns. Specifically, the spatial correlation matrix of the noise is estimated during noise-only segments, and the steering vectors of the beamformer are estimated during single active speaker segments. The beamformer's weights are not updated during concurrent activities of more than one speaker.    
In \cite{yousefi2021real, kanda2020joint}, both \ac{CNN} and attention mechanisms are employed for speaker-counting and identification. 
In \cite{8462548, 9053096}, a \ac{LSTM} model is used for the task of \ac{OSD}. Unlike the \ac{CSD}, only two classes comprise the \ac{OSD} task. The first comprises noise-only or single-speaker segments, while the second comprises overlapped speech segments, namely two or more concurrent speakers.

The Transformer model, which was originally proposed in the \ac{NLP} domain \cite{9222960, vaswani2017attention}, was later adopted by the audio community for various tasks, e.g., speech separation \cite{9413901} and audio classification \cite{gong21b_interspeech}.
It was demonstrated in \cite{gong21b_interspeech} that the \ac{AST} model, which is an adaptation of the \ac{ViT} model \cite{dosovitskiy2020image}, outperforms \ac{CNN}-based models. 
We stress that the \ac{AST} model only processes single-microphone data, whereas multiple microphones are available in many real-world use cases. 
It is well-known that, if properly utilized, the additional spatial information may improve performance.  
In \cite{cornell:hal-02908241}, a model based on \ac{TCN} is used, and in \cite{CORNELL2022101306}, a Transformer-based model is used. Both models estimate the activity of the speakers. Specifically, two related tasks are implemented, \ac{VAD} and \ac{OSD}, as well as their joint estimation. The \ac{VAD} classifies audio segments into two classes: 1) no active speaker and 2) speech activity (either one or more speakers). We can therefore refer to the combined \ac{VAD} and \ac{OSD} task as a \ac{CSD} task.
A Transformer-based solution is also utilized for audio \ac{OSD} \cite{zheng2021beamtransformer} and for audio-visual \ac{OSD} \cite{10064301}.
The \ac{CSD}, which is a multi-class classification task, is more complex than \ac{OSD} or \ac{VAD}, which are binary classification tasks.

In the current contribution, we propose an algorithm to solve the \ac{CSD} task. Our contribution is threefold: 1) we extend the use of \ac{ViT} and adapt it to the multi-microphone case, 2) we incorporate in the training process a re-weighting mechanism according to the importance of each class and further use calibration to improve the classification accuracy, and finally 3) similarly to \cite{cornell:hal-02908241, CORNELL2022101306, 10064301, 8462548, zheng2021beamtransformer, 9053096}, we evaluate the performance of the proposed model on AMI \cite{10.1007/11677482_3} and CHiME~5 \cite{barker2018is} databases, and additionally on the recently introduced AliMeeting \cite{Yu2022M2MeT} database (in Chinese).
 
\section{Problem Formulation}
\label{Problem Formulation}
Let $X_i(\ell,k),\,i=1,\ldots,N$ represent the \ac{STFT} of the microphone signals, where $N$ is the number of microphones, $\ell$ and $k$ represent the frame index and the frequency index, respectively.
The goal of a \ac{CSD} algorithm is to classify each audio segment (either single-microphone or multi-microphone) into one of the three classes: 
\begin{equation} \label{eq: classes}
\mathrm{CSD}(\ell) = 
\begin{cases}
 \textrm{Class \#0} &  \textrm{Noise only}  \\
 \textrm{Class \#1} &  \textrm{Single-speaker activity}   \\ 
 \textrm{Class \#2} &  \textrm{Concurrent-speaker activity} 
\end{cases}.
\end{equation} 
The statistical characteristics of the audio segments may change according to the scenario. Multiple types of noise may exist for class `0' (`Noise-Only'). Class `1' (`Single-speaker activity') can be challenging due to the variability of human speech. Individuals may have different accents, speaking styles, and vocal characteristics, making it difficult for algorithms to identify them accurately. In class `2' (`Concurrent-speaker activity'), the different number of active speakers may result in diverse statistical properties.
In addition, the presence of background noise or reverberation can further complicate the task.
Consequently, developing robust and accurate \ac{CSD} methods that can handle a wide range of input conditions becomes essential.

\section{Proposed Model}
\label{Proposed Model}
The proposed \ac{CSD} model is based on the \ac{ViT} \cite{dosovitskiy2020image} architecture, which has achieved state-of-the-art performance on a variety of computer vision tasks. We have modified the original \ac{ViT} architecture to better suit audio processing requirements, including the use of log-spectrum as input and the ability to handle both single-channel and multichannel audio.
The input features are the log-magnitude of the \ac{STFT} of the audio signals, denoted hereinafter log-spectrum.

The model consists of three main blocks: Embedding, Transformer, and Classification.
The first block linearly projects the input data and generates the input tokens for the Transformer model. The second block is a \ac{MHA} transformer block, consisting of self-attention layers, which can capture complex relations within its input data. The attention mechanism \cite{vaswani2017attention} allows the model to simultaneously focus on different parts of the input, enabling it to learn rich feature representations from the log-spectrum. Finally, several fully-connected layers are applied to map the learned features to the final output predictions.

\begin{figure*}[htbp]
    \centering
      \includegraphics[width=0.99\textwidth]{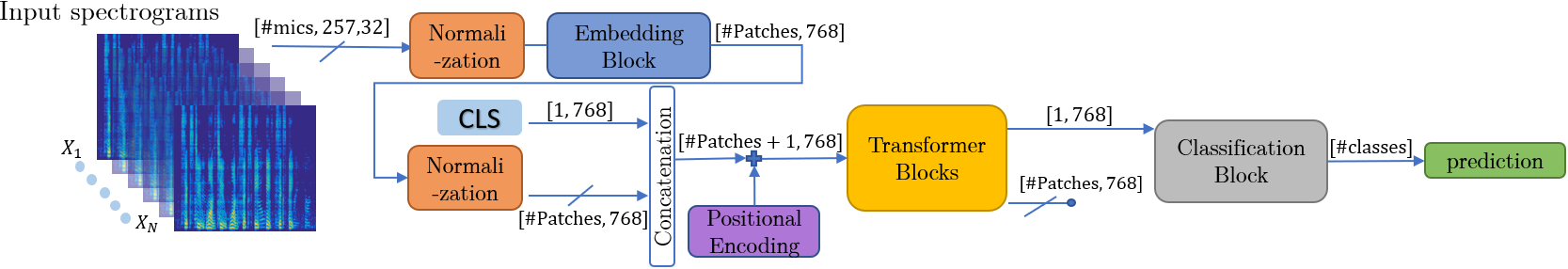}
      \setlength{\belowcaptionskip}{0pt}
      \caption{High-level architecture of the proposed model.}
      \label{model_arch_high_level}
   \end{figure*}

Our starting point is, therefore, the \ac{ViT} model, with the images substituted by the log-spectra and the RGB channels by the multi-microphone measurements. The multichannel model attends to different areas in the input to achieve the best classification results. 
We use a \ac{CE} loss function for training, a common choice for classification tasks. In addition, we used \ac{LS} \cite{muller2019does} and the \acf{CS} loss function \cite{galdran2020cost} as regularization techniques to improve the ability of the model to generalize to unknown data.
The high-level architecture of the model is presented in Fig.~\ref{model_arch_high_level}.

\subsection{Pre-Processing and Input Features}
\label{Features}
The microphone signals are first resampled to 16kHz (if the original sampling rate differs from the nominal one). Subsequently, the signals are analyzed by an \ac{STFT} with Hann window of length 512 and 50\% overlap. Finally, the log-spectrum is calculated, resulting in 257 frequency bins per frame. 

The output labels are determined using the transcribed databases, with a resolution of 0.1~Sec and context frames 0.2~Sec long on both sides of the analyzed segment. Therefore, each audio segment, lasting 0.5 seconds, is categorized into one of the three classes.
The overall dimensions of the input tensor are $N \times 257 \times 32$, where $N$ is the number of microphones, $257$ is the number of frequency bins, and $32$ is the number of time bins.

\subsection{Architecture}
\label{Architecture}
We will now elaborate on the three blocks comprising the model's architecture.

The Embedding block receives the input log spectrum and produces the tokens utilized by the Transformer block. This block splits the input data into patches and linearly projects each to form tokens. We set the dimensions of a 2-D learnable kernel and strides values in the \ac{TF} axes, such that each patch is projected to an embedding space with a dimension $D$ and considered a token. This process results in a features tensor with dimensions $\#\textrm{Tokens}\times D$.
This block can be designed to handle both multichannel and single-channel signals, as will be explained below. Following \cite{kumar2023dual}, we apply a dual patch-normalization for improving the \ac{ViT} results (see the `Normalization' block in Fig.~\ref{model_arch_high_level}).

The Transformer block follows the \ac{ViT} architecture with minor modifications. The Transformer was initialized with random weights and trained with the chosen databases. This module consists of several layers of \ac{MHA} blocks. We followed the \ac{ViT} architecture and used Class token [CLS], an additional learnable token added as an input to the Transformer. In addition, a learnable positional embedding is added to each input token before the first \ac{MHA} layer. In total, the input and output of each \ac{MHA} blocks are tokens of shape $(\#\textrm{Tokens}+1)\times D$.

The last block in our model, the Classification block, consists of two fully connected layers that map the Transformer output to fit the number of classes. The Classification block takes only the token corresponding to [CLS] as input. This should make the classification process unbiased towards any particular token, as discussed in \cite{dosovitskiy2020image}.

\subsection{Single- and multichannel Embedding Blocks}
\label{Single-Channel to multichannel embedding block}
The Embedding block transforms the input data into tokens used by the Transformer block, comprising 12 Transformers. We stress that the AST model \cite{gong21b_interspeech} addresses a different classification task and is limited to single-channel inputs. Since we are also interested in the multichannel case, the standard embedding block should be modified accordingly.

\begin{figure*}
     \centering
     \begin{subfigure}[b]{0.49\textwidth}
         \centering
\includegraphics[width=0.97\textwidth]{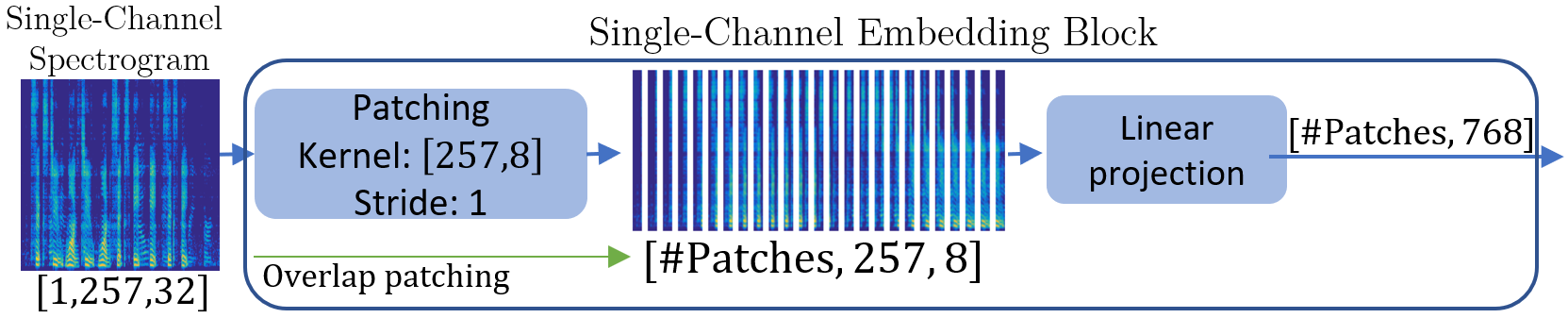}
      \caption{Single-Channel Embedding Block: Embedding layer for a single-channel log-spectrum.}
      \label{embed_single_channel}
     \end{subfigure}
     \hfill
     \begin{subfigure}[b]{0.49\textwidth}
         \centering
\includegraphics[width=0.97\textwidth]{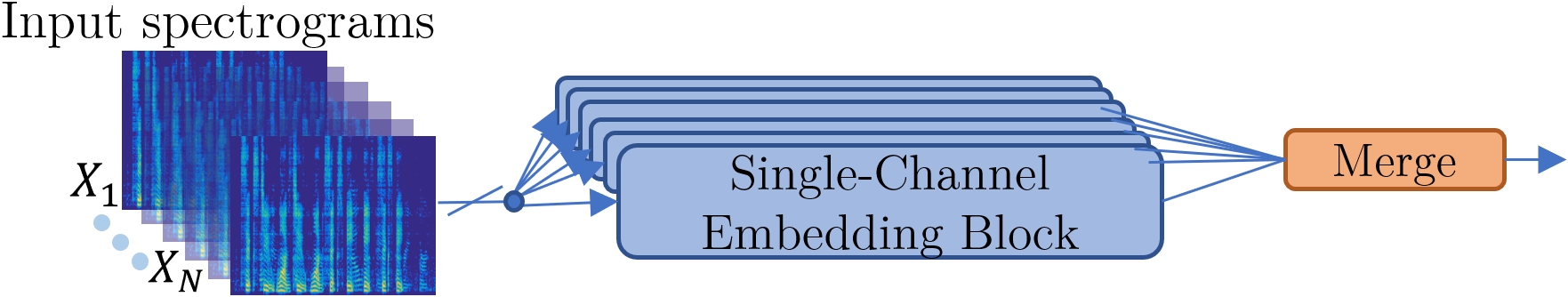}
      \caption{multichannel Embedding Block: 
      The `merge' operation stands for either summation, averaging, or concatenation. All `single channel blocks' weights can be either shared or distinct.}
      \label{embed_multi_channel}
      \end{subfigure}
      
      \caption{Embedding Block: Details.}
\end{figure*}

The basic, single-microphone structure, i.e.~$N=1$, is depicted in Fig.~\ref{embed_single_channel}. The input log spectrum is split into patches with a 2-D learnable kernel with a stride set to 1. 
In \ac{ViT}, the shape of the patches is $16\times 16$, but according to our analysis, a more useful patch size is $257\times 8$, as it jointly analyzes the entire frequency axis. Later on, each patch is linearly projected to a dimension of $D=768$, resulting in a tensor of shape $\#\textrm{Patches}\times 768$.

The information must be merged from the different channels for the multichannel case. The overall proposed structure is depicted in Fig.~\ref{embed_multi_channel}.
The most effective merging technique, denoted here as Type \#1, entails independently applying a single-channel embedding to each microphone signal and then combining all channels through concatenation. This process yields an output tensor with dimensions $N\cdot\#\textrm{Patches}\times 768$. Ensuring identical channel numbers during both the training and testing stages is necessary for this structure. Furthermore, due to the expansion of the channel dimension compared to the single-microphone scenario, there is an increase in the number of input tokens for the subsequent Transformer block. Similarly, the input feature vector to the Classification block also increases. All of which increase the total number of parameters.

To further analyze the merging strategies, we have examined two alternatives, designated hereinafter Type \#2 and Type \#3.

In Type \#2, each single-channel embedding block is independently applied, and a summation operation merges the information. The resulting data shape is $\#\textrm{Patches}\times 768$. While the independent processing of each channel may be beneficial performance-wise, it also requires the number of channels to be identical in the training and test stages.

In Type \#3, the weights of all single-channel embedding blocks are shared (Siamese networks), and their output is then merged using an averaging operation. The shape of the result is again $\#\textrm{Patches}\times 768$. This structure is indifferent to a mismatch between the number of microphones in the training and test stages and simultaneously reduces the number of parameters. Nevertheless, it may fall short of fully capturing the relationships between the signals from the microphones.

After experimenting with all three alternatives, we decided to use Type \#1 due to its ability to perform cross-channel attention, a property that enhances the overall performance of the proposed method. Table~\ref{tab: Merging block comparison} compares the \ac{mAP} results for the three merging types for all databases. Table~\ref{tab: Merging block comparison} further supports choosing Type \#1 since it outperforms all three types for the \ac{OSD} task while exhibiting only marginal performance degradation for the \ac{VAD} task.

\begin{table}[htbp]
    \centering
    \caption{Ablation study for merging strategies: The \acf{mAP} (\%) measure for the \ac{VAD} and the \ac{OSD} classifiers, as well as the number of required parameters (\#P) in millions (M).}
    \label{tab: Merging block comparison}
\resizebox{\columnwidth}{!}{%
\begin{tabular}{lccccccccc}\toprule
& \multicolumn{3}{c}{AMI} & \multicolumn{3}{c}{AliMeeting} & \multicolumn{3}{c}{CHiME}
\\\cmidrule(lr){2-4}\cmidrule(lr){5-7}\cmidrule(lr){8-10}
              & \ac{VAD}  & \ac{OSD} & \#P (M)   & \ac{VAD}  & \ac{OSD} & \#P (M) &  \ac{VAD} & \ac{OSD} & \#P (M)
           \\\midrule
Type \#1   & 98  & \textbf{73.1}   & 98.1       & 98.2  & \textbf{87.8} & 98.1M      & 91.6  & \textbf{83.5}  & 91.8 \\
Type \#2   & \textbf{98.1} & 69.5  & 98.1       & \textbf{99.6} & 73.8  & 98.1M      & \textbf{96.6} & 56.5   & 91.7 \\
Type \#3   & 97.9  & 71.9          & 86.9       & 98.3  & 86.4          & 86.9M      & 91.9          & 63.4   & 86.9
\\\bottomrule
\end{tabular}
}
\end{table}

%
%
%
%
%
\subsection{Objective Functions}
\label{Objectives}
As we aim at the classification task, the natural choice for a loss function is the \acf{CE}.
However, the classification results were not balanced between the different classes in our databases. We have, therefore, used two additional methods to guide the model to focus on the more challenging examples in the data, mainly Class \#1. We used \acf{LS} \cite{muller2019does} and class weights.\footnote{\texttt{https://towardsdatascience.com/class-weights-\\for-categorical-loss-1a4c79818c2d}} Both demonstrated an improvement in classification accuracy.
We also applied the \acf{CS} loss \cite{galdran2020cost}. In this procedure, we follow a 2-stage training procedure: First, we train the model with no \ac{CS} loss, then we modify the \ac{CS} loss weights according to the results and retrain the model. The \ac{CS} loss weights are defined as a $3\times 3$ matrix that gives more weight, i.e., increasing the loss function, for more frequent classification errors.
When applying the \ac{CS} loss, we use an extra hyper-parameter weighting between the \ac{CS} loss and the \ac{CE} loss. This hyper-parameter was tested extensively and was set to a value of 15-20 (depending on the different databases).
\subsection{Confidence Calibration}
\label{Confidence-Calibration}
After training the model, we applied confidence calibration using temperature scaling \cite{pmlr-v70-guo17a}. The model calibration can become handy when the model is used to control the estimation of the building blocks of a beamformer \cite{8553564}, where erroneous classification can significantly deteriorate the separation performance. Calibration enhances the interpretability of the model's predictions by aligning them with probabilities, establishing an appropriate threshold for considering the prediction valid. An instance of utilizing a calibrated model involves setting segments with predictions below a certain threshold to Class \#2. These segments are then excluded from the estimation of the beamformer's weights.
\section{Experimental study}
\label{Experimental study}
\subsection{Databases}
We applied the proposed model to three real-world databases, namely AMI \cite{10.1007/11677482_3}, AliMeeting \cite{Yu2022M2MeT}, and CHiME 5 \cite{barker2018is}. All databases use a microphone array for the recordings, AMI and AliMeeting use an 8-microphone array, and CHiME 5 uses a 4-microphone array. AMI database consists of 100 hours of meeting recordings with English speakers (both male and female) in 3 different rooms and setups. The AliMeeting comprises 118.75 hours of real meetings with 2-4 participants speaking in Mandarin.
The CHiME 5 database consists of recordings of conversations between English speakers from different real-home environments. CHiME~5 has six arrays (U01 to U06) with four microphones each. We chose to train and report only for arrays U01 and U02.

All databases are fully transcribed with phrase-level resolution. This allows us to generate ground-truth labels for training the model.
The distribution of the different classes is depicted in Table~\ref{tab: Class frequency}, where it is evident that the classes are unbalanced. The phrases are dominated by Class \#1, as typical to natural human conversations. This imbalance between the different classes must be addressed during training to avoid biased classification results. This is done by modifying the cost function during training, as elaborated above.
\begin{table}[htbp]
    \caption{Class frequency [\%] in all databases.}
    \label{tab: Class frequency}
    \centering
\begin{tabular}{ p{2.8cm}ccc  }
 \toprule
 Database/Class& \#0 & \#1& \#2\\
  \midrule
 Ali-Meeting    & 6.9     & 67.2   & 25.9\\
 AMI            & 16.8    & 71.8   & 11.4\\
 CHiME 5        & 20.5    & 50.9   & 28.6\\
 \bottomrule
\end{tabular}
\end{table}
We split the databases into train, validation, and test sets. 

\noindent\textbf{CHiME~5 database:}\footnote{According to the official website, CHiME~5 and CHiME~6 databases are identical, and only the challenge description differs.} We used the existing Train-Val-Test split given on the challenge’s official website. Note that we chose to remove all utterances with saturated sound files. We demonstrated our results using only U01 and U02, while \cite{CORNELL2022101306} uses all arrays U01-U06 and presents the average results.

\noindent\textbf{AMI\footnote{\texttt{groups.inf.ed.ac.uk/ami/corpus/datasets.shtml}} and AliMeeting\footnote{\texttt{www.openslr.org/119/}} databases:} We used the official Train-Val-Test splits as provided in the databases' documentation. 

%
\subsection{Algorithm Setup}
\label{Setup}
 We used the architecture described in Section~\ref{Single-Channel to multichannel embedding block} with Type \#1 Embedding block and \ac{CS} loss for all models since, as discussed above, it stands out as the most effective scheme. In training the models, we used the Adam optimizer with a learning rate of $1e^{-6}$, a weight decay of $1e^{-9}$, and a batch size of 128. To prevent overfitting, considering the model's substantial parameter count, we limit the number of epochs to a range of 10-15, depending on the examined database. The overall parameter count falls within the range of 86.9-98.1M, as illustrated in Table~\ref{tab: Merging block comparison}.
We used the following hyperparameters: The Embedding block is set with a dimension of $D=768$, and the Transformer block is set with 12 heads and a depth of 12. The classification block has one hidden layer with dimension 387.

\subsection{Results}
\label{Results}
Choosing the right evaluation metric is essential because our model should classify between three classes. 
In this work, especially when compared to competing methods, we will evaluate the performance of the proposed scheme in terms of Precision, Recall, and \acf{mAP}. 
To gain further insights into the proposed method's performance and give a more detailed analysis of the errors, we also chose to present the confusion matrix that compares the ground-truth labels with the predicted labels by our model (as a percentage normalized to the ground-truth labels). 
The confusion matrices for the single-microphone case are depicted in Table~\ref{tab: Single-Channel results} and Table~\ref{tab: multichannel results} for the multi-microphone case. 
\begin{table}[ht]
    \centering
    \caption{\ac{CSD} results: Single-microphone model confusion matrices, as [\%] normalized to the ground-truth labels. `T'-true labels, `P'-predicted labels.}
    \label{tab: Single-Channel results}
\begin{tabular}{lccccccccc}\toprule
& \multicolumn{3}{c}{AMI} & \multicolumn{3}{c}{AliMeeting} & \multicolumn{3}{c}{CHiME}
\\\cmidrule(lr){2-4}\cmidrule(lr){5-7}\cmidrule(lr){8-10}
           T \textbackslash P & 0  & 1 & 2   & 0  & 1 & 2 &  0 & 1 & 2
           \\\midrule
0        & 78 & 20 & 2     & 88 & 11 & 1      & 55 & 32 & 13 \\
1        & 9  & 75 & 16    & 10 & 77 & 13     & 10 & 51 & 39 \\
2        & 1  & 37 & 62    & 2  & 30 & 68     & 3  & 34 & 63
\\\bottomrule
\end{tabular}
\vspace{-.2cm}
\end{table}

\begin{table}[ht]
    \centering
    \caption{\ac{CSD} results: Multi-microphone model confusion matrices, as [\%] normalized to the ground-truth labels. Using the Type \#1 embedding block. `T'-true labels, `P'-predicted labels.}
    \label{tab: multichannel results}
\begin{tabular}{lccccccccc}\toprule
& \multicolumn{3}{c}{AMI} & \multicolumn{3}{c}{AliMeeting} & \multicolumn{3}{c}{CHiME}
\\\cmidrule(lr){2-4}\cmidrule(lr){5-7}\cmidrule(lr){8-10}
           T \textbackslash P & 0  & 1 & 2   & 0  & 1 & 2 & 0  & 1 & 2
           \\\midrule
0        & 80 & 18 & 2    & 85 & 13 & 2     & 73 & 21 & 6 \\
1        & 10 & 74 & 16   & 7  & 82 & 11    & 19 & 51 & 30\\
2        & 2  & 35 & 63   & 2  & 28 & 70    & 8  & 33 & 59
\\\bottomrule
\end{tabular}
\vspace{-.4cm}
\end{table}
%
%
%
%
%

A performance comparison, in terms of the Precision, Recall, and \ac{mAP} (\%) metrics for the \ac{OSD} task for the AMI database is depicted in Table~\ref{tab: AMI Database Comparison}.
We compare both our single- and multichannel variants to several algorithms from the literature \cite{CORNELL2022101306,10064301,9053096,zheng2021beamtransformer, bredin2021end}. It is worth noting that \cite{10064301} introduces three models employing distinct modalities for the \ac{OSD} task: 1) audio-only, 2) video-only, and 3) audiovisual. Our reference pertains to the audio-only model. We emphasize that our multi-class \ac{CSD} task is inherently more intricate than the binary \ac{OSD} task. We transformed our \ac{CSD} classification results into two binary classification tasks to facilitate comparison. This involved combining two classes: for the \ac{VAD} task, classes \#1 and \#2 were aggregated, while for the \ac{OSD} task, classes \#0 and \#1 were aggregated. Notably, the proposed model demonstrates a significant performance superiority over competing methods in the \ac{OSD} task.

As outlined in Table~\ref{tab: multichannel model mAP comparison}, a comparison between the proposed multi-microphone model and \cite{CORNELL2022101306} using the \ac{mAP} metric reveals a significant improvement in the \ac{OSD} task. However, there is a slight decrease in performance for the \ac{VAD} task. The AliMeeting database stands out with the most favorable classification results, while the CHiME database proves to be the most challenging among the three tested databases.
\begin{table}[htb]
    \caption{A comparison between the proposed single- and multi-microphone variants and various competing methods in evaluating the performance on the \ac{OSD} task, including Precision, Recall, and \ac{mAP} (\%) measures on the AMI Database.}
    \label{tab: AMI Database Comparison}
    \centering
\begin{tabular}{ p{1.8cm}cccc }
 \toprule
 Variant & Method & Precision & Recall & \ac{mAP} \\
  \midrule
   & \cite{CORNELL2022101306} & N/A & N/A & 59.1 \\
              \multirow{3}{0pt}{Single-channel}              & \cite{10064301} & N/A & N/A & 62.7 \\
                            & \cite{9053096}  & 86.8 & 65.8   & N/A \\
                            & pyannote 2.0 \cite{bredin2021end}  & 80.7 & 70.5   & N/A \\
                            & \textbf{Our}  & \textbf{91.4} & \textbf{88.9} & \textbf{69.3} \\
 \midrule
     & \cite{zheng2021beamtransformer} & 87.8  & 87  & N/A \\
                      \multirow{2}{0pt}{multichannel}     & \cite{CORNELL2022101306} & 87.8  & 87 & 60.3 \\
                            & \textbf{Our} & \textbf{92.4} & \textbf{89} & \textbf{73.1}\\
 \bottomrule
\end{tabular}
\end{table}
\vspace{-.2cm}
\begin{table}[htbp]
    \centering
    \caption{Multi-microphone model: Comparison of \ac{mAP} (\%)  for \ac{VAD} and \ac{OSD} tasks, tested over all three databases.}
    \label{tab: multichannel model mAP comparison}
\begin{tabular}{lccccccccc}\toprule
& \multicolumn{2}{c}{\ac{VAD}} & \multicolumn{2}{c}{\ac{OSD}}
\\\cmidrule(lr){2-3}\cmidrule(lr){4-5}
           & \cite{CORNELL2022101306}  & Ours   & \cite{CORNELL2022101306} & Ours
           \\\midrule
AMI         & \textbf{98.7}  & \textbf{98.7}     & 60.3        &\textbf{73.1}  \\
CHiME       & \textbf{95.4}  & 91.6              & 52.4        &\textbf{83.5}  \\
AliMeeting  & N/A            & 98.2              & N/A         & 87.8  \\
\bottomrule
\end{tabular}
\vspace{-11pt}
\end{table}

\section{Conclusions}
\label{Conclusions}
In this paper, we introduced a multi-microphone transformer-based model designed for the \ac{CSD} task, accompanied by a training scheme capable of assigning weights to classes based on their importance and incorporating a calibration stage. Through experiments, we illustrated the practicality of our proposed model in real-world databases, showcasing its performance advantages compared to existing methods. Notably, the model exhibits versatility, proving effective in both single- and multi-microphone scenarios, with a distinct advantage observed in the latter.

 \balance
\bibliographystyle{IEEEtran}
\bibliography{my_library}

\begin{thebibliography}{10}
\providecommand{\url}[1]{#1}
\def\UrlFont{\rmfamily}
\providecommand{\newblock}{\relax}
\providecommand{\bibinfo}[2]{#2}
\providecommand\BIBentrySTDinterwordspacing{\spaceskip=0pt\relax}
\providecommand\BIBentryALTinterwordstretchfactor{4}
\providecommand\BIBentryALTinterwordspacing{\spaceskip=\fontdimen2\font plus
\BIBentryALTinterwordstretchfactor\fontdimen3\font minus
  \fontdimen4\font\relax}
\providecommand\BIBforeignlanguage[2]{{%
\expandafter\ifx\csname l@#1\endcsname\relax
\typeout{** WARNING: IEEEtran.bst: No hyphenation pattern has been}%
\typeout{** loaded for the language `#1'. Using the pattern for}%
\typeout{** the default language instead.}%
\else
\language=\csname l@#1\endcsname
\fi
#2}}

\bibitem{8553564}
S.~E. Chazan, J.~Goldberger, and S.~Gannot, ``{LCMV} beamformer with
  {DNN}-based multichannel concurrent speakers detector,'' in \emph{26th
  European Signal Processing Conference (EUSIPCO)}, 2018, pp. 1562--1566.

\bibitem{yousefi2021real}
M.~Yousefi and J.~H. Hansen, ``{Real-Time Speaker Counting in a Cocktail Party
  Scenario Using Attention-Guided Convolutional Neural Network},'' in
  \emph{Proc. Interspeech 2021}, 2021, pp. 1484--1488.

\bibitem{kanda2020joint}
N.~Kanda, Y.~Gaur, \emph{et~al.}, ``{Joint Speaker Counting, Speech
  Recognition, and Speaker Identification for Overlapped Speech of any Number
  of Speakers},'' in \emph{Proc. Interspeech 2020}, 2020, pp. 36--40.

\bibitem{8462548}
N.~Sajjan, S.~Ganesh, \emph{et~al.}, ``Leveraging lstm models for overlap
  detection in multi-party meetings,'' in \emph{IEEE International Conference
  on Acoustics, Speech and Signal Processing (ICASSP)}, 2018, pp. 5249--5253.

\bibitem{9053096}
L.~Bullock, H.~Bredin, and L.~P. Garcia-Perera, ``Overlap-aware diarization:
  Resegmentation using neural end-to-end overlapped speech detection,'' in
  \emph{ICASSP 2020 - 2020 IEEE International Conference on Acoustics, Speech
  and Signal Processing (ICASSP)}, 2020, pp. 7114--7118.

\bibitem{9222960}
A.~Gillioz, J.~Casas, \emph{et~al.}, ``Overview of the transformer-based models
  for {NLP} tasks,'' in \emph{15th Conference on Computer Science and
  Information Systems (FedCSIS)}, 2020, pp. 179--183.

\bibitem{vaswani2017attention}
A.~Vaswani, N.~Shazeer, \emph{et~al.}, ``Attention is all you need,''
  \emph{Advances in neural information processing systems (NeurIPS)}, vol.~30,
  2017.

\bibitem{9413901}
C.~Subakan, M.~Ravanelli, \emph{et~al.}, ``Attention is all you need in speech
  separation,'' in \emph{IEEE International Conference on Acoustics, Speech and
  Signal Processing (ICASSP)}, 2021, pp. 21--25.

\bibitem{gong21b_interspeech}
Y.~Gong, Y.-A. Chung, and J.~Glass, ``{AST: Audio Spectrogram Transformer},''
  in \emph{Proc. Interspeech}, 2021, pp. 571--575.

\bibitem{dosovitskiy2020image}
A.~Dosovitskiy, L.~Beyer, \emph{et~al.}, ``An image is worth 16x16 words:
  Transformers for image recognition at scale,'' in \emph{International
  Conference on Learning Representations (ICLR)}, 2021.

\bibitem{cornell:hal-02908241}
S.~Cornell, M.~Omologo, \emph{et~al.}, ``{Detecting and counting overlapping
  speakers in distant speech scenarios},'' in \emph{Proc. Interspeech},
  Shanghai, China, Oct. 2020.

\bibitem{CORNELL2022101306}
------, ``Overlapped speech detection and speaker counting using distant
  microphone arrays,'' \emph{Computer Speech \& Language}, vol.~72, p. 101306,
  2022.

\bibitem{zheng2021beamtransformer}
S.~Zheng, S.~Zhang, \emph{et~al.}, ``Beamtransformer: Microphone array-based
  overlapping speech detection,'' \emph{arXiv preprint arXiv:2109.04049}, 2021.

\bibitem{10064301}
M.~Kyoung, H.~Jeon, and K.~Park, ``Audio-visual overlapped speech detection for
  spontaneous distant speech,'' \emph{IEEE Access}, vol.~11, pp.
  27\,426--27\,432, 2023.

\bibitem{10.1007/11677482_3}
J.~Carletta, S.~Ashby, \emph{et~al.}, \emph{Machine Learning for Multimodal
  Interaction}.\hskip 1em plus 0.5em minus 0.4em\relax Springer Berlin
  Heidelberg, 2006, ch. The {AMI} Meeting Corpus: A Pre-announcement, pp.
  28--39.

\bibitem{barker2018is}
J.~Barker, S.~Watanabe, \emph{et~al.}, ``The fifth `chime’ speech separation
  and recognition challenge: Dataset, task and baselines,'' in
  \emph{Proceedings Interspeech}, Hyderabad, India, Sept. 2018.

\bibitem{Yu2022M2MeT}
F.~Yu, S.~Zhang, \emph{et~al.}, ``M2{M}e{T}: The {ICASSP} 2022 multi-channel
  multi-party meeting transcription challenge,'' in \emph{IEEE International
  Conference on Acoustics, Speech and Signal Processing (ICASSP)}, 2022.

\bibitem{muller2019does}
R.~M{\"u}ller, S.~Kornblith, and G.~E. Hinton, ``When does label smoothing
  help?'' \emph{Advances in neural information processing systems (NeurIPS)},
  vol.~32, 2019.

\bibitem{galdran2020cost}
A.~Galdran, J.~Dolz, \emph{et~al.}, ``Cost-sensitive regularization for
  diabetic retinopathy grading from eye fundus images,'' in \emph{Medical Image
  Computing and Computer Assisted Intervention (MICCAI)}, 2020, pp. 665--674.

\bibitem{kumar2023dual}
M.~Kumar, M.~Dehghani, and N.~Houlsby, ``Dual {PatchNorm},'' \emph{Transactions
  on Machine Learning Research}, 2023.

\bibitem{pmlr-v70-guo17a}
C.~Guo, G.~Pleiss, \emph{et~al.}, ``On calibration of modern neural networks,''
  in \emph{Proceedings of the 34th International Conference on Machine
  Learning}, vol.~70, Aug. 2017, pp. 1321--1330.

\bibitem{bredin2021end}
H.~Bredin and A.~Laurent, ``End-to-end speaker segmentation for overlap-aware
  resegmentation,'' \emph{arXiv preprint arXiv:2104.04045}, 2021.

\end{thebibliography}

\end{document}